\documentclass[copyright,creativecommons]{eptcs}
\providecommand{\event}{ACL2 2013}
\usepackage{breakurl}
\usepackage{xspace}
\usepackage{graphicx}

\usepackage[english]{babel}
\usepackage[utf8]{inputenc}
\usepackage{times}
\usepackage[T1]{fontenc}
\usepackage{lipsum}
\usepackage{ragged2e}
\usepackage{alltt}
\usepackage{rotating}
\usepackage{lscape}
\usepackage{setspace}
\usepackage{array}
\usepackage{verbatim}
\usepackage{fancyvrb}
\usepackage{upquote}
\usepackage{color}

\begin{document}
\title{Abstract Stobjs and \\ Their Application to ISA Modeling}

\author{
    Shilpi Goel \qquad\qquad Warren A Hunt, Jr. \qquad\qquad Matt Kaufmann
    \institute{Department of Computer Science, University of Texas at Austin}
    \email{shigoel@cs.utexas.edu \qquad\qquad hunt@cs.utexas.edu \qquad\qquad kaufmann@cs.utexas.edu}
}

\def\titlerunning{Abstract Stobjs for ISA Modeling}
\def\authorrunning{S. Goel, W. Hunt, and M. Kaufmann}

\maketitle

\begin{abstract}

We introduce a new ACL2 feature, the {\em abstract stobj}, and show how
to apply it to modeling the instruction set architecture of a
microprocessor.  Benefits of abstract stobjs over traditional
(``concrete'') stobjs can include faster execution, support for
symbolic simulation, more efficient reasoning, and resilience of proof
developments under modeling optimization.

\end{abstract}




\section{Introduction}

In support of our modeling and verification efforts for
microprocessors, we have introduced a new ACL2 event to support the
definition of {\em abstract stobjs}.  The traditional single-threaded
objects supported by ACL2, ``concrete'' {\em stobjs}~\cite{stobjs},
are well known to support efficient execution.  While they
allow a user to specify datatype restrictions for each defined field,
they do not permit restrictions involving more than one field.  Such
restrictions can be necessary for defining an {\em invariant} that
specifies the allowable states for a stobj.  Of course, we can define
a predicate that specifies the relationships between the fields of the
stobj for this purpose.  However, such a predicate may be expensive to
execute during guard checking, difficult to prove during guard
verification, and complicate theorems by cluttering up the
hypotheses, thereby making these theorems hard to use as well.

An abstract stobj can solve these and other problems by providing an
alternative logical interface to a previously-defined concrete stobj.
When introducing an abstract stobj, we prove once and for all that it
remains in ``lockstep'' correspondence with its associated concrete
stobj. Thus, the user can define a simpler logical representation of
the concrete stobj in order to abstract away its complexity for
reasoning.

The goal of this paper is to introduce abstract stobjs to the ACL2
community so that ACL2 users can consider using this feature in their
proof developments.  Thus we begin, in Section~\ref{sec-abs-stobj}, by
outlining abstract stobjs and working a very simple example.  Then in
Section~\ref{reasoning-on-proc-models}, we illustrate how to take
advantage of abstract stobjs for a more realistic sort of application:
modeling a microprocessor and reasoning about programs running on it.
We conclude with a discussion of the benefits provided by abstract
stobjs.

Those who wish to use abstract stobjs in their own work may find it
useful to consult the documentation topic for
\texttt{defabsstobj}~\cite{defabsstobj-doc}.  Those interested in
going below the user level are, of course, welcome to peruse the
source code; in particular, the logical foundations are sketched in a
long comment~\cite{essay-absstobjs}.

\section{Abstract Stobjs}~\label{sec-abs-stobj}

The development of ACL2 has been guided by a desire for ACL2 programs to
execute efficiently.  A typical performance issue for functional languages is
that when using list data structures, read and write operations are linear in
the length of the list.  Tree-like structures can help, but still require
consing for writes, which can be expensive.  Thus, ACL2 has long supported {\em
single-threaded objects}, or {\em stobjs}, which are mutable
objects with applicative semantics.

ACL2 Version 5.0 introduced a related feature, {\em abstract stobjs}.
Let us refer to (ordinary) stobjs as ``concrete stobjs.''  Just as
concrete stobjs are introduced with the \texttt{defstobj} event,
abstract stobjs are introduced with the \texttt{defabsstobj} event.
In this section, we explain abstract stobjs at a high level and then
illustrate their use with a simple pedagogical example.  We conclude
by discussing an atomicity issue that can arise, together with a
discussion of how one can deal with it.

\subsection{Abstract stobjs in the abstract}~\label{sec-abs-stobj-abs}

An abstract stobj may be viewed as an alternative representation of a
{\em corresponding} concrete stobj, where the abstract stobj
recognizer may impose an invariant that specifies additional
requirements.  An abstract stobj is accessed (for reading, writing, or
both) by defining {\em exports}: functions whose logical (or {\em
  abstract}) function is established by the \texttt{:LOGIC} keyword,
which is what ACL2 reasons about; and whose executable (or {\em
  concrete}) function is specified by the \texttt{:EXEC} keyword,
and is what ACL2 actually executes when applied to the new stobj.
The concrete functions, which were earlier introduced to operate on
the concrete stobj, now also operate on the abstract stobj, which is a
raw Lisp structure that is produced by a new call of the concrete
stobj's creator function in raw Lisp.
That is, the raw Lisp abstract and concrete stobjs are instances of
the same data structure but are distinct, with no shared structure;
and concrete stobj primitives execute on both the concrete and the
abstract stobj in raw Lisp.

A \texttt{defabsstobj} event specifies a {\em correspondence
  predicate}.  A proof obligation ensures preservation of this
predicate upon update of the abstract stobj, in the spirit of
bisimulation, as illustrated by the commutative diagram below.  Assume
that a \texttt{defabsstobj} event has introduced an abstract stobj
\verb|st|, a corresponding concrete stobj \verb|st$c|, and a function
\verb|f| associated with \texttt{:LOGIC} and \texttt{:EXEC} functions
\verb|f$a| and \verb|f$c| that update the abstract and concrete stobj,
respectively.  Then the diagram below states that \verb|st$c1|
corresponds to \verb|st1| provided that the following hypotheses hold.

\begin{itemize}

\item \verb|f$a| maps instance \verb|st0| of \verb|st| to
  \verb|st1|.

\item \verb|f$c| maps instance \verb|st$c0| of \verb|st$c| to
  \verb|st$c1|.

\item The correspondence predicate holds for \verb|st$c0| and \verb|st0|.

\end{itemize}




\begin{center}
\includegraphics[width=3.5in]{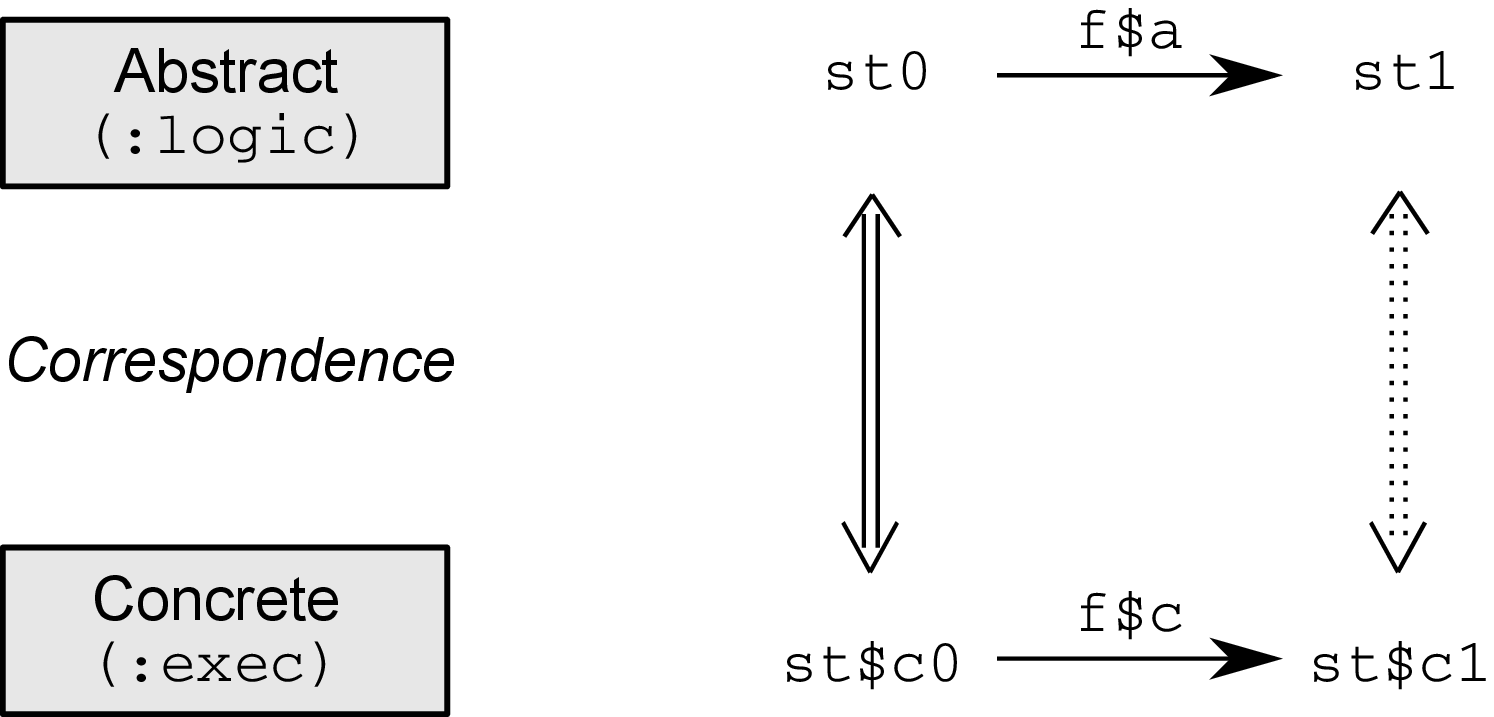}
\label{fig:commutative-diagram}
\end{center}

A \texttt{defabsstobj} event specifies a recognizer, a creator, and
exports.  For each exported function \verb|f|, a \texttt{:LOGIC} ({\em
  abstract}) function is specified that is logically equal to
\verb|f|, and an \texttt{:EXEC} ({\em concrete}) function is specified
that operates on the corresponding concrete stobj.  All of these
functions must be defined before a \texttt{defabsstobj} event is
evaluated, as this event generates proof obligations about these
functions.  The proof obligations are represented as events, which
ACL2 must admit before the \texttt{defabsstobj} event is admitted.
But the generated events will probably not all go through
automatically, in which case ACL2 prints out those that remain to be
proved, so that the user can formulate and prove necessary lemmas in
advance.  In summary, a \texttt{defabsstobj} event will typically be
introduced as follows.

\begin{enumerate}

\item Introduce a concrete stobj using \texttt{defstobj}.

\item Define all \texttt{:LOGIC} and \texttt{:EXEC} functions.  (Of course,
  \texttt{:EXEC} functions that are primitives, introduced in the step
  above, need not be defined again here.)

\item Define the correspondence predicate.

\item Prove the required events that are printed upon evaluation of the
  \texttt{defabsstobj} event.

\item Admit the \texttt{defabsstobj} event.

\end{enumerate}

\subsection{An example}\label{abstract-stobjs-example}

We illustrate abstract stobjs using an example.  We give only some
highlights below; for full details, see the supporting
materials~\cite{absstobj-support}.


We begin by defining a concrete stobj, with two fields: a memory of
100 natural number values (initially 100 zeroes), and a
``miscellaneous'' (``misc'') field that can contain an arbitrary
value.

{\footnotesize
\begin{Verbatim}[commandchars=\\\{\}]
(defstobj st$c
  (mem$c :type (array t (100)) :initially 0)
  misc$c)
\end{Verbatim}
}

The next step is to define all the \texttt{:LOGIC} functions for our
abstract stobj.  We begin with the recognizer, \verb|st$ap|.  In our
simple example, it is convenient to think of two ``fields'' that
correspond to those of the above concrete stobj, but to make things
interesting, we use an entirely different data structure for our abstract
stobj than for our concrete stobj: here, a \texttt{cons} whose
\texttt{car} is arbitrary (for ``misc'') and whose \texttt{cdr}
corresponds to the memory.

The following recursive function recognizes the implementation of
memory for our abstract stobj.  Unlike the memory of our concrete
stobj, this memory is based on an association list.  Just for fun, we
add an invariant beyond what is required of the concrete stobj: all
memory values are even natural numbers.

{\footnotesize
\begin{Verbatim}[commandchars=\\\{\}]
(defun mem-map$ap (x)
  (declare (xargs :guard t))
  (cond ((atom x) (null x))
        ((atom (car x)) nil)
        (t (and (natp (caar x)) (< (caar x) 100) \textsl{; index is in range}
                (natp (cdar x)) (evenp (cdar x)) \textsl{; value is an even natural number}
                (mem-map$ap (cdr x))))))
\end{Verbatim}
}

\noindent Now we can define the \texttt{:LOGIC} functions for our
abstract stobj recognizer and creator.

{\footnotesize
\begin{Verbatim}[commandchars=\\\{\}]
(defun st$ap (x)
  (declare (xargs :guard t))
  (and (consp x)
       (mem-map$ap (cdr x))))

(defun create-st$a ()
  (declare (xargs :guard t))
  (cons nil nil)) \textsl{; (cons misc mem)}
\end{Verbatim}
}

\noindent We choose exported functions that read and write the ``misc'' and
memory of our abstract stobj.

{\footnotesize
\begin{Verbatim}[commandchars=\\\{\}]
(defun misc$a (st$a)
  (declare (xargs :guard (st$ap st$a)))
  (car st$a))

(defun update-misc$a (v st$a)
  (declare (xargs :guard (st$ap st$a)))
  (cons v (cdr st$a)))

(defun lookup$a (k st$a)
  (declare (xargs :guard (and (natp k) (< k 100)
                              (st$ap st$a))))
  (let* ((mem-map (cdr st$a))
         (pair (assoc k mem-map)))
    (if pair (cdr pair) 0)))

(defun update$a (k val st$a)
  (declare (xargs :guard (and (st$ap st$a)
                              (natp k) (< k 100)
                              (natp val) (evenp val))))
  (cons (car st$a)
        (put-assoc k val (cdr st$a))))
\end{Verbatim}
}

Our next task is to define the correspondence function \verb|st$corr|,
which relates concrete and abstract stobj instances.  Since this
relation is of logical interest only, we avoid guards and guard
verification.

{\footnotesize
\begin{Verbatim}[commandchars=\\\{\}]
(defun corr-mem (n st$c st$a) \textsl{; auxiliary to st$corr, defined below}
  (declare (xargs :stobjs st$c :verify-guards nil))
  (cond ((zp n) t)
        (t (let ((i (1- n)))
             (and (equal (mem$ci i st$c) (lookup$a i st$a))
                  (corr-mem i st$c st$a))))))

(defun st$corr (st$c st$a)
  (declare (xargs :stobjs st$c :verify-guards nil))
  (and (st$cp st$c)
       (st$ap st$a)
       (equal (misc$c st$c) (misc$a st$a))
       (corr-mem 100 st$c st$a)))
\end{Verbatim}
}

We are ready to evaluate our \texttt{defabsstobj} event --- not to
admit it yet, but to print events to the terminal that we paste into
the book under development.

{\footnotesize
\begin{Verbatim}[commandchars=\\\{\}]
(DEFABSSTOBJ ST
  :EXPORTS ((LOOKUP :EXEC MEM$CI)
            (UPDATE :EXEC UPDATE-MEM$CI)
            MISC UPDATE-MISC))
\end{Verbatim}
}

\noindent The events printed out partition naturally into three
classes, according to the three suffixes used: \\ 
\verb|{CORRESPONDENCE}|, \verb|{PRESERVED}|, and \verb|{GUARD-THM}|.
We consider these in turn.  For brevity, we ignore events pertaining to
the ``misc'' field.

The first \verb|{CORRESPONDENCE}| theorem below guarantees that
initial concrete and abstract stobjs correspond.  The second says that
for exported function \texttt{LOOKUP}, the \texttt{:EXEC} and \texttt{:LOGIC}
functions applied to corresponding states produce the same value.  The
third corresponds to the commutative diagram discussed above: for
exported function \texttt{UPDATE}, the \texttt{:EXEC} and \texttt{:LOGIC}
functions applied to corresponding states produce corresponding
states.

{\footnotesize
\begin{verbatim}
(DEFTHM CREATE-ST{CORRESPONDENCE}
        (ST$CORR (CREATE-ST$C) (CREATE-ST$A))
        :RULE-CLASSES NIL)

(DEFTHM LOOKUP{CORRESPONDENCE}
        (IMPLIES (AND (ST$CORR ST$C ST)
                      (NATP I) (< I 100)
                      (ST$AP ST))
                 (EQUAL (MEM$CI I ST$C)
                        (LOOKUP$A I ST)))
        :RULE-CLASSES NIL)

(DEFTHM UPDATE{CORRESPONDENCE}
        (IMPLIES (AND (ST$CORR ST$C ST)
                      (ST$AP ST)
                      (NATP I) (< I 100)
                      (NATP V) (EVENP V))
                 (ST$CORR (UPDATE-MEM$CI I V ST$C)
                          (UPDATE$A I V ST)))
        :RULE-CLASSES NIL)
\end{verbatim}
}

The \verb|{PRESERVED}| theorems guarantee that the recognizer always
holds for our abstract stobj; it holds initially, and it is preserved
by any well-guarded application of \verb|UPDATE|.  There cannot be
such a preservation theorem for \verb|LOOKUP|, because it does not
return a new value of the abstract stobj, \verb|ST|.  Preservation of
the recognizer justifies an optimization: an abstract stobj recognizer
is defined for execution (in raw Lisp) to return \texttt{T} when
applied to a stobj object (an array, in raw Lisp).  Since that
recognizer can be defined logically as an arbitrarily complex
invariant, this is an important optimization.  We say more about how
recognizer evaluation benefits execution in Section~\ref{conclusion}.

{\footnotesize
\begin{verbatim}
(DEFTHM CREATE-ST{PRESERVED}
        (ST$AP (CREATE-ST$A))
        :RULE-CLASSES NIL)

(DEFTHM UPDATE{PRESERVED}
        (IMPLIES (AND (ST$AP ST)
                      (NATP I) (< I 100)
                      (NATP V) (EVENP V))
                 (ST$AP (UPDATE$A I V ST)))
        :RULE-CLASSES NIL)
\end{verbatim}
}

To see the significance of the \verb|{GUARD-THM}| theorems below,
consider an ill-guarded call on argument list $(i_0,v_0,\texttt{st})$
of the function \texttt{update}, introduced by the
\texttt{defabsstobj} above.
A guard violation occurs, even when
guard-checking has been turned off, in which case an error message
says that ``ACL2 does not support non-compliant
live stobj manipulation.'' This is because ACL2 {\em always} checks
the guards of functions applied to stobjs, for functions introduced by
the \texttt{defabsstobj} event, and thus a corresponding call of
\texttt{update\$c} is made on
argument list $(i_0,v_0,\texttt{st\$c})$ only if the guard of the
original call of \texttt{update} was satisfied. The \verb|{GUARD-THM}|
for \texttt{update} states that the guard must therefore be satisfied
for the call of \texttt{update\$c}, which ensures ``compliant live
stobj manipulation''.

{\footnotesize
\begin{Verbatim}[commandchars=\\\{\}]
(DEFTHM LOOKUP\{GUARD-THM\} ... ){\em ; omitted to save space}
\end{Verbatim}

\begin{verbatim}
(DEFTHM UPDATE{GUARD-THM}
        (IMPLIES (AND (ST$CORR ST$C ST)
                      (ST$AP ST)
                      (NATP I) (< I 100)
                      (NATP V) (EVENP V))
                 (AND (INTEGERP I)
                      (<= 0 I)
                      (< I (MEM$C-LENGTH ST$C))))
        :RULE-CLASSES NIL)
\end{verbatim}
}

Now we are ready to submit our \texttt{defabsstobj} event.  We present it
in a more verbose form than given above, in order to illustrate
default naming conventions.  The few parts retained from the short
form above are in CAPITAL LETTERS; the rest simply fills in defaults.

{\footnotesize
\begin{Verbatim}[commandchars=\\\{\}]
(DEFABSSTOBJ ST
  :concrete st$c \textsl{; the corresponding concrete stobj}
  :recognizer (stp :logic st$ap :exec st$cp)
  :creator (create-st :logic create-st$a :exec create-st$c
                      :correspondence create-st\{correspondence\}
                      :preserved create-st\{preserved\})
  :corr-fn st$corr \textsl{; a correspondence function (st$corr st$c st)}
  :EXPORTS ((LOOKUP :logic lookup$a
                    :EXEC MEM$CI
                    :correspondence lookup\{correspondence\}
                    :guard-thm lookup\{guard-thm\})
            (UPDATE :logic update$a
                    :EXEC UPDATE-MEM$CI
                    :correspondence update\{correspondence\}
                    :preserved update\{preserved\}
                    :guard-thm update\{guard-thm\})
            (MISC :logic misc$a
                  :exec misc$c
                  :correspondence misc\{correspondence\})
            (UPDATE-MISC :logic update-misc$a
                         :exec update-misc$c
                         :correspondence update-misc\{correspondence\}
                         :preserved update-misc\{preserved\})))
\end{Verbatim}
}

A \texttt{defabsstobj} event gives its exports signatures that enforce
single-threadedness.  However, the logical functions retain their
original signatures.  For example, the function \verb|misc| introduced
above takes a stobj, \texttt{st}, as an argument; but function
\verb|misc$a| continues to take an ordinary argument, which presents
no problems since subsequent stobj-based code would be written using
\verb|misc|, not \verb|misc$a|.

\subsection{An atomicity issue}

We conclude our overview by explaining an issue that may arise if one
decides to use abstract stobjs.  In short, the correctness of abstract
stobjs relies on preservation of recognizers, which can be at risk due
to non-atomic updates by exported functions.  Note that this problem
does not arise with concrete stobjs, since a \texttt{defstobj} event
introduces functions that update atomically.

Our initial implementation of \texttt{defabsstobj} in ACL2 Version 5.0
had a soundness bug, as illustrated by the following events based on
the bug report from Sol Swords.
Note that the abstract stobj is updated by an exported function that
logically makes more than one call of the concrete stobj's updater
functions, but that sequence of calls doesn't complete.  The resulting
state then violates the abstract stobj recognizer.  We say that such
exported functions are not {\em atomic}.

{\footnotesize
\begin{Verbatim}[commandchars=\\\{\}]
(defstobj const-stobj$c (const-fld$c :initially 0))

(defstub stop () nil)

(defun change-fld$c (const-stobj$c) \textsl{; Logically, this sets the field to 0.}
  (declare (xargs :stobjs const-stobj$c))
  (let ((const-stobj$c (update-const-fld$c 1 const-stobj$c)))
    (prog2$ (stop) \textsl{; aborts, leaving the field at value 1}
            (update-const-fld$c 0 const-stobj$c))))

(defun const-stobj$ap (const-stobj$a)
  (declare (xargs :guard t))
  (equal const-stobj$a 0))

(defun z (const-stobj$a)
  (declare (xargs :guard t) (ignore const-stobj$a))
  0)

(defun create-const-stobj$a ()
  (declare (xargs :guard t))
  0)

(defun-nx const-stobj-corr (const-stobj$c const-stobj$a)
  (equal const-stobj$c '(0)))

(in-theory (disable (const-stobj-corr) (change-fld$c)))

\textsl{; Events generated by defabsstobj would go here but are not shown.}

(defabsstobj const-stobj
  :concrete const-stobj$c
  :recognizer (const-stobjp :logic const-stobj$ap :exec const-stobj$cp)
  :creator (create-const-stobj :logic create-const-stobj$a
                               :exec create-const-stobj$c)
  :corr-fn const-stobj-corr
  :exports ((get-fld :logic z :exec const-fld$c)
            (change-fld :logic z :exec change-fld$c)))
\end{Verbatim}
}

\noindent In ACL2 Version 5.0 we can see a violation of the logical definition
of \texttt{get-fld} as \texttt{z}.

{\footnotesize
\begin{verbatim}
  ACL2 !>(change-fld const-stobj)


  ACL2 Error in TOP-LEVEL:  ACL2 cannot ev the call of undefined function
  STOP on argument list:

  NIL

  To debug see :DOC print-gv, see :DOC trace, and see :DOC wet.

  ACL2 !>(get-fld const-stobj)
  1
  ACL2 !>
\end{verbatim}
}

\noindent In ACL2 Version 6.0, however, the above \texttt{defabsstobj}
event fails with the following error message.

{\footnotesize
\begin{verbatim}
  ACL2 Error in ( DEFABSSTOBJ CONST-STOBJ ...):  The :EXEC field CHANGE-FLD$C,
  specified for defabsstobj field CHANGE-FLD, appears capable of modifying
  the concrete stobj, CONST-STOBJ$C, non-atomically; yet :PROTECT T was
  not specified for this field.  See :DOC defabsstobj.
\end{verbatim}
}

As suggested by the message, one is now required to specify
\texttt{:PROTECT T} in a \texttt{defabsstobj} for any exported
function that might not execute to completion.  Fortunately, ACL2
applies some syntactic analysis to detect exported functions that are
atomic --- that is, invoke at most one updater call for the
corresponding concrete stobj --- and these do not need the
\texttt{:PROTECT} keyword.  In Section~\ref{model-abs}, we see that
this keyword argument is only needed for one export of abstract stobj
\texttt{x86-32}.

ACL2 generates extra code for an exported function marked with
\texttt{:PROTECT T}, to support a check that atomicity has not been
violated.  That check is made at the top level and also when
completing book certification.  When the check fails (rarely, in our
experience), an error occurs, and book certification is disabled for
the remainder of the session in order to prevent unsoundness.  Why
does ACL2 not simply eliminate the error?  For one, there is no way in
general to roll back to a state in which the abstract stobj recognizer
holds, since the \texttt{:EXEC} function could make arbitrary changes
to the abstract stobj before being interrupted.  Of course, ACL2 could
simply reinitialize the abstract stobj; but we suspect that users
would prefer to manage this situation themselves.

A debug mode is available that provides a more informative error
message, indicating which update operation was incomplete.  Although
the debug mode is not terribly slow, nevertheless efficiency is a key
goal for stobj (and abstract stobj) execution, so the debug mode is
off by default.

\section{Reasoning on Processor Models}~\label{reasoning-on-proc-models}

In this section we will show how abstract stobjs can benefit the
development and use of a processor model whose state is modeled with a
stobj.  Our model employs an interpreter approach to operational
semantics~\cite{mechanized-op-semantics} that is routinely used to
formalize models in ACL2.  We start by reviewing that approach.

\subsection{Interpreter Approach to Operational Semantics}

ACL2 has been successfully used to formalize a number of ISA models
using a classic interpreter approach to operational semantics. There
are four main components in a model formalized using this approach; we
describe these in the context of our Y86
model~\cite{bryant2001computer}, which is a very simple 32-bit microprocessor
model that has an X86-like ISA.

\begin{itemize}

\item {\em State}: We define the {\em state} of the processor to
  contain registers and the memory address space.  For the sake of
  execution efficiency in the case of the Y86, we model the state with
  stobjs.

\item {\em Instruction Semantic Functions}: We give semantics to each
  instruction by defining a function that takes the machine state and
  returns the modified state. This {\em instruction semantic function}
  describes the effect of executing the instruction by modifying the
  processor state.

\item {\em Step Function}: We then define a {\em step} function that
  executes a single instruction.  This function fetches the
  instruction from the memory, decodes it, and then dispatches control
  to the semantic function corresponding to that instruction.

\item {\em Run Function}: Finally, we define the {\em run} function,
  which calls the {\em step} function repeatedly until the program runs
  to completion, the number of instructions to be run becomes zero, or
  an error occurs. This {\em run} function specifies the processor
  model.

\end{itemize}

For more details about the basic Y86 model in ACL2, see ACL2 community book
directory \\
\texttt{models/y86/y86-basic/}.

\subsection{Y86 ISA Model without Abstract Stobjs}~\label{y86-two-level}

A space-efficient memory model is important when modeling real
processors (which, in the case of a contemporary processor, can have a
memory of up to $2^{52}$ bytes, i.e.,
4096 terabytes), in order to keep the memory footprint of the model
manageable.  Hunt and Kaufmann~\cite{hunt-kaufmann-fmcad-2012}
implemented a formal processor model which has space-efficient memory
as well as high-speed performance.  Here we adapt that model to the
Y86.  For details, see ACL2 community book directory
\texttt{models/y86/y86-two-level/}.

{\footnotesize
\begin{Verbatim}[commandchars=\\\{\}]
(defstobj x86-32$c
  ...
  \textsl{;; the program counter}
  (eip$c :type (unsigned-byte 32)
         :initially 0)
  ...
  \textsl{;; the memory model: space-efficient implementation}
  (mem-table :type (array (unsigned-byte 32)
                          (*mem-table-size*)) \textsl{;; *mem-table-size* = 256}
             :initially 1
             :resizable nil)
  (mem-array :type (array (unsigned-byte 8)
                          (*initial-mem-array-length*)) \textsl{;; 1,677,721,600}
             :initially 0
             :resizable t)
  (mem-array-next-addr :type (integer 0 4294967296)
                       :initially 0)
  ...
  :renaming ((x86-32$cp x86-32$cp-pre))
  )
\end{Verbatim}
}

We define the state of the processor, \texttt{x86-32\$c}, to contain
registers and memory address space.  There are three
memory-related fields: \texttt{mem-table}, \texttt{mem-array}, and
\texttt{mem-array-next-addr}.  Note that the stobj recognizer has
been renamed to \texttt{x86-32\$cp-pre}.

The basic idea behind the memory model is simple --- memory is
allocated on demand instead of all at once.  Memory is implemented as
a flat array of fixed-size consecutive blocks (16MB blocks here).
\texttt{mem-table} stores the addresses of blocks (or rather, the
addresses for the first byte of each
block),\\ \texttt{mem-array-next-addr} stores the address of the block
to be allocated next, and \texttt{mem-array} is the real memory where
bytes are stored.  Hence, we think of an address of a byte in the
memory (i.e., index of \texttt{mem-array}) to be composed of two parts
--- the address of the block and the offset within the block.

This stobj definition requires us to maintain a stronger invariant on
the processor state than the stobj recognizer \texttt{x86-32\$cp-pre},
which merely says that all the fields are well-formed.  The stronger
recognizer should also assert that the relationship among the three
memory fields gives a well-formed memory.  We call this recognizer
\texttt{x86-32\$cp}.

{\footnotesize
\begin{Verbatim}[commandchars=\\\{\}]
(defun x86-32$cp (x86-32$c)
  (declare (xargs :stobjs x86-32$c))
  (and (x86-32$cp-pre x86-32$c)
       (good-memp x86-32$c))) \textsl{;; Complicated predicate!}
\end{Verbatim}
}

The memory write function \texttt{!mem\$ci} for \texttt{x86-32\$c} is
as follows. Note that it reads one field of the stobj,
\texttt{mem-table}, then potentially re-sizes another field --- \texttt{mem-array} ---
based on the value read earlier (i.e., a value in \texttt{mem-table}),
and finally updates \texttt{mem-array} appropriately.

{\footnotesize
\begin{Verbatim}[commandchars=\\\{\}]
(defun !mem$ci (i v x86-32$c)
  (declare (xargs :stobjs x86-32$c \textsl{;; enforces syntactic restriction on stobjs}
                  :guard (and (integerp i) (<= 0 i) (< i *mem-size-in-bytes*)
                              (n08p v)
                              (x86-32$cp x86-32$c)))) \textsl{;; enforces} \texttt{good-memp}
  (let* ((i-top (ash i -24))
         (addr (mem-tablei i-top x86-32$c)))
    (mv-let (addr x86-32$c)
            (cond ((eql addr 1) \textsl{ ;; Page is not present.}
                   (add-page-x86-32$c i-top x86-32$c)) \textsl{ ;; potential resizing}
                  (t (mv addr x86-32$c)))
            (!mem-arrayi (logior addr (logand #xffffff i)) v x86-32$c))))
\end{Verbatim}
}

{\em Reasoning about Y86 Programs}

Though such a definition of the processor state goes a long way
towards obtaining execution efficiency, it presents some problems for
reasoning.

In this section, we focus on one such problem: impediments to using
the GL package~\cite{swords2011verified}. GL is a framework for
proving ACL2 theorems involving finite objects; it uses symbolic
execution as a proof procedure. The reason we choose to use GL is that
we hope to prove snippets of code in large programs correct fully
automatically using GL's ability to compute with symbolic objects.

As a starting point, we will attempt to reason about a very simple
program.

{\footnotesize
\begin{Verbatim}[commandchars=\\\{\}]
(defconst *simple-program-source*
  '(                   \textsl{;; Main program}
    (pos   80)         \textsl{; 80: Align to 16-byte address}
    main
    (irmovl  1023 %eax)
    halt-of-main
    (halt)             \textsl{; 86: Halt}
    end-of-code        \textsl{; 87: Label for the end of the code}
    (pos 8192)         \textsl{; 8192: Assemble position; "stack" has value 8192}
    stack))
\end{Verbatim}
}

We wish to prove, via GL's symbolic execution, that the register
\texttt{\%eax} has value 1023 and the instruction pointer points to
the halt address 86 at the end of this program.  The stobj creator
function \texttt{create-x86-32\$c} gives us a symbolic ACL2 object
corresponding to the processor state \texttt{x86-32\$c}.  However,
since it can not be used directly in functions, we can define a
state-initializing non-executable function as follows:

{\footnotesize
\begin{Verbatim}[commandchars=\\\{\}]
(defun-nx simple-program-init-x86-32$c (eip)
  (declare (xargs :guard (n32p eip)))
  (init-y86-state
   nil                       \textsl{; Y86 status}
   eip                       \textsl{; Initial program counter}
   nil                       \textsl{; Initial stack pointer}
   nil                       \textsl{; Initial flags, if NIL, then all zeros}
   *simple-program-binary*   \textsl{; Initial memory}
   (create-x86-32$c))        \textsl{; Create the processor state}
  )
\end{Verbatim}
}

To verify the guards of \texttt{simple-program-init-x86-32\$c}
painlessly, it is prudent to prove:

{\footnotesize
\begin{Verbatim}[commandchars=\\\{\}]
(defthm x86-32$cp-create-x86-32$c
  (x86-32$cp (create-x86-32$c)))
\end{Verbatim}
}

We wish to prove this theorem by taking advantage of ACL2's ability to
reason by evaluating terms without free variables, so that we can
avoid the effort of formulating suitable lemmas.  Unfortunately, as
ACL2 tries to prove this theorem it calls \texttt{create-x86-32\$c},
which prevents the proof from completing because of the attempt to
create a \texttt{mem-array} list of length 1,677,721,600.  Our
solution is to introduce a single lemma to be proved by computation on
the raw Lisp stobj.  Note that logically, \texttt{with-local-stobj}
generates a call of the stobj creator function for its first argument.

{\footnotesize
\begin{Verbatim}[commandchars=\\\{\}]
(defun hack ()
  (with-local-stobj x86-32$c
                    (mv-let (result x86-32$c)
                            (mv (x86-32$cp x86-32$c) x86-32$c)
                            result)))

(defthm x86-32$cp-create-x86-32$c
  (x86-32$cp (create-x86-32$c))
  :hints (("Goal" :use (hack)
           :in-theory (union-theories '((hack)) (theory 'minimal-theory)))))
\end{Verbatim}
}

Finally, we try to prove correctness using the
\texttt{def-gl-thm} macro provided by the GL package.

{\footnotesize
\begin{Verbatim}[commandchars=\\\{\}]
(def-gl-thm y86-simple-program-correct
  :hyp (equal esp 8192)
  :concl (let* ((start-eip (cdr (assoc-eq 'main
                                          *simple-program-symbol-table*)))
                (halt-eip (cdr (assoc-eq 'halt-of-main
                                         *simple-program-symbol-table*)))
                \textsl{;; Initialize the x86-32 state.}
                (x86-32$c (simple-program-init-x86-32$c start-eip))
                (count 300)
                \textsl{;; Run the processor for}\texttt{ count }\textsl{steps.}
                (x86-32$c (y86 x86-32$c count)))
           (and (equal (rgfi *mr-eax* x86-32$c)
                       1023)
                (equal (eip x86-32$c)
                       halt-eip)))
  :g-bindings `((esp (:g-number ,(gl-int 0 1 15)))))
\end{Verbatim}
}

As GL complains about the clock running out, we increase the clock by
adding \texttt{:concl-clk \\10000000000000000} to the
\texttt{def-gl-thm}. Now, however, there is a value stack overflow.

GL does symbolic execution according to logical definitions of ACL2
functions, so it does not provide stobj performance. As the logical
representation of a stobj is a linear list of its fields --- which,
for arrays, can themselves be linear lists --- large lists have to be
created in order to symbolically execute functions that take the state
as input. For this model, the \texttt{mem-array} list is so large that
merely creating it results in a stack overflow, let alone
accessing/updating it using linear traversals.

Can we somehow avoid the stack overflow?  One approach might seem to
be to change the way we use the GL package, so that it can handle such
functions better. For example, we can define a GL clause processor
that will allow \texttt{make-list-ac} (the list creator function) to
execute directly on concrete values instead of being interpreted. Even
that will not be of much help in this situation because the lists are
too large.  A second idea could be to change the implementation of
some GL functions in order to make them more efficient --- however,
they are memoized and hence not something we can make tail
recursive~\cite{sswords} to get higher performance.  Yet a third idea
could be to do proofs for simple programs using a
\texttt{with-local-stobj} technique similar to what we have used above
for the proof of \texttt{x86-32\$cp-create-x86-32\$c}: define a
function like the \texttt{hack} (above) that would return \texttt{T}
if our post-condition holds. However, this is possible only with
concrete data, not an arbitrary 32-bit input.

Of course, reasoning about code using GL, or indeed any tool that uses
bit-blasting for symbolic execution, is bound to hit limits for models
with large arrays.  The challenge is then to find a path for
proceeding when that happens.  We now see how abstract stobjs provide
such a path.

\subsection{Y86 ISA  Model with Abstract Stobjs}\label{model-abs}

A small processor state would be amenable to proof by symbolic
execution.  We can define an abstract stobj over the concrete stobj to
obtain such a state.  The memory field in the abstract stobj is
defined using a sparse data structure, a
record~\cite{Kaufmann02efficientrewriting}, which is a finite
normalized structure that associates non-default values to keys.  The
initial representation of the abstract memory field is now
\texttt{nil}, as opposed to a large linear list of zeroes for the
concrete memory field.  The abstract memory contains only those values
that have been written to the memory explicitly.  We describe this
approach below; for details, see ACL2 community book directory
\texttt{models/y86/y86-two-level-abs/}.

Our abstract memory field corresponds to the functionality provided by
the three concrete memory fields. The following definition suffices
for the recognizer of the abstract memory field:

{\footnotesize
\begin{Verbatim}[commandchars=\\\{\}]
(defun-sk memp (x)
  (forall i
          (implies (g i x) \textsl{ ;;}\texttt{ g }\textsl{is a record `get' function}
                   (and (n32p i)
                        (n08p (g i x))))))
\end{Verbatim}
}

\noindent The \texttt{:LOGIC} definition of the memory write function \texttt{!mem\$ai} is
as follows:
{\footnotesize
\begin{Verbatim}[commandchars=\\\{\}]
(defun !mem$ai (i v x86-32)
  (declare (xargs :guard (and (x86-32$ap x86-32)
                              (n32p i)
                              (n08p v))))
  (update-nth *memi*
              (s i v (nth *memi* x86-32))
              x86-32))
\end{Verbatim}
}

\noindent Note that it is a considerably simpler definition than
\texttt{!mem\$ci}.\\

Here is the abstract stobj definition:

{\footnotesize
\begin{Verbatim}[commandchars=\\\{\}]
(defabsstobj x86-32
  :concrete x86-32$c
  :recognizer (x86-32p :logic x86-32$ap :exec x86-32$cp-pre)
  :creator (create-x86-32 :logic create-x86-32$a :exec create-x86-32$c)
  :corr-fn corr
  :exports (...
            (eip :logic eip$a :exec eip$c)
            (!eip :logic !eip$a :exec !eip$c)
            ...
            \textsl{;; !mem$ci is our complicated memory write function.}
            (!memi :logic !mem$ai :exec !mem$ci :protect t)))
\end{Verbatim}
}

The recognizer \texttt{x86-32\$ap} is similar to
\texttt{x86-32\$cp-pre}, except that the three memory field
recognizers have been replaced by \texttt{memp}.  Similarly, the
creator \texttt{create-x86-32\$a} is similar to
\texttt{create-x86-32\$c} except for \texttt{nil} being the initial
memory instead of linear lists for the three (logical) memory fields.  The
correspondence function states that every field apart from the memory
fields of the concrete and abstract stobjs is the same and the memory
fields correspond as follows:

{\footnotesize
\begin{Verbatim}[commandchars=\\\{\}]
(defun-sk corr-mem (x86-32$c abs-mem-field)
  \textsl{;; Looking up an address in the memory of the concrete stobj returns the}
  \textsl{;; same value as looking it up in the memory of the abstract stobj.}
  (forall i
          (implies (and (natp i)
                        (< i *mem-size-in-bytes*))
                   (equal (mem$ci i x86-32$c)
                          \textsl{;; next line is (or (g i abs-mem-field) 0))}
                          (g0 i abs-mem-field)))))

\end{Verbatim}
}

{\em Reasoning about Y86 Programs}

Proving theorems using GL's symbolic execution is significantly more
viable for the Y86 model with abstract stobjs, because the abstraction
provides a smaller representation of the processor state and simpler
logic definitions of memory read and write functions.  We also note
that proving \texttt{(x86-32p (create-x86-32))} (by execution) without
the \texttt{with-local-stobj} technique is no longer prohibitive for
this model, again because of the smaller state representation.

In the supporting materials~\cite{absstobj-support}, we define a constant
\texttt{*popcount-source*} whose value represents a program which
counts the number of ones (`on' bits) in its input, written in the Y86
assembly language.  We have proved a correctness property of this
program for the model with abstract stobjs, using GL's symbolic
execution.  Note that we did this without first proving any lemmas or
defining any additional GL clause processor. The time taken to prove
this theorem was \textasciitilde29s on a 2.2 GHz Intel Core i7 Apple
MacBook Pro with a memory of 8GB, running ACL2 Version 6.0 built on
Clozure Common Lisp.

{\footnotesize
\begin{Verbatim}[commandchars=\\\{\}]
(def-gl-thm y86-popcount-correct
  :hyp (and (equal esp 8192)
            \textsl{;; n, a 32-bit unsigned integer, is the input.}
            (n32p n))
  :concl (let* ((start-eip (cdr (assoc-eq 'call-popcount *popcount-symbol-table*)))
                (halt-eip (cdr (assoc-eq 'halt-of-main *popcount-symbol-table*)))
                \textsl{;; Initialize the x86-32 state.}
                (x86-32 (popcount-init-x86-32 n esp start-eip))
                (count 300)
                \textsl{;; Run the processor count times}
                (x86-32 (y86 x86-32 count)))
           \textsl{;; At the end of the run, the eax register will have}
           \textsl{;; the logcount of the input n and the instruction }
           \textsl{;; pointer will be at the halt instruction.}
           (and (equal (rgfi *mr-eax* x86-32) (logcount n))
                (equal (eip x86-32) halt-eip)))
  :g-bindings `((n   (:g-number ,(gl-int 0 2 33)))
                (esp (:g-number ,(gl-int 1 2 15))))
  :rule-classes nil)
\end{Verbatim}
}

\noindent Compare this to our failed attempt in Subsection~\ref{y86-two-level} to
prove a program as simple as \\\texttt{*simple-program-source*} correct
on the model without abstract stobjs.

\section{Conclusion}\label{conclusion}

We saw that incorporating an abstract stobj into a model entails a
significant amount of work --- the logic versions of the fields'
accessor and updater functions, the stobj creator function, and the
recognizer functions have to be defined, a correspondence function has
to be provided, and finally, the proof obligations (preservation,
correspondence, and guard theorems) have to be met.  However, our Y86
example suggests that the benefits of using abstract stobjs can
outweigh the requisite effort.  For more realistic models than the
Y86, the benefits can be even more significant.  We now discuss some
benefits of using abstract stobjs.

\begin{itemize}
\item {\em Execution in the ACL2 loop:}\\
\texttt{x86-32\$cp}, an expensive predicate, appears in the guard of
the functions that have \texttt{x86-32\$c} as an input.  For example,
the run function of the Y86 model without abstract stobjs has the
following \texttt{declare} statement:

{\footnotesize
\begin{Verbatim}[commandchars=\\\{\}]
(declare (xargs :guard (and (natp n) (x86-32$cp x86-32$c))
                :stobjs (x86-32$c)))
\end{Verbatim}
}

When such a function is executed on concrete data in the ACL2 loop,
execution is slow because guard-checking is costly.  However, for the
analogous run function that takes abstract stobj \texttt{x86-32} as
input, execution on concrete data in the ACL2 loop does not suffer
from this expensive guard check.  Here is the declare statement of the
run function of the Y86 model that uses an abstract stobj:

{\footnotesize
\begin{Verbatim}[commandchars=\\\{\}]
(declare (xargs :guard (natp n) :stobjs (x86-32)))
\end{Verbatim}
}

As mentioned in Section~\ref{abstract-stobjs-example}, calls of
abstract stobj recognizer functions trivially evaluate to \texttt{T},
taking advantage of the fact that the recognizer always holds.  This
observation explains why \texttt{memp} could be safely defined as a
non-executable function, even though it supports the logical
definition of recognizer \texttt{x86-32p} (see
Section~\ref{model-abs}).

\item {\em Symbolic Execution using GL}:\\
In the previous section, we saw how abstract stobjs made symbolic
execution using GL feasible.
We are using abstract stobjs to great benefit in our
X86 modeling (which is much more complicated than our Y86
modeling).  We have used GL to do code proofs of real
X86 binaries~\cite{goel-hunt-popcount}.  Of course, we do not claim
that we can prove all programs correct using symbolic
execution. However, having such a capability certainly reduces the
proof development time. We can use GL for proving parts of a large
program correct and then use traditional theorem proving
techniques~\cite{raymoore} to compose these proofs to obtain a proof
of correctness of the entire program.

\item {\em Simplifying reasoning:}\\
Reasoning about functions that take \texttt{x86-32\$c} as input
involves proving the hypotheses of invariance theorems.  For example,
the memory read-over-write theorem is:

{\footnotesize
\begin{Verbatim}[commandchars=\\\{\}]
(defthm read-write
  (implies (and (x86-32$cp x86-32$c)
                (integerp i) (<= 0 i) (< i *mem-size-in-bytes*)
                (integerp j)
                (<= 0 j) (< j *mem-size-in-bytes*)
                (n08p v))
           (equal (memi j (!mem$ci i v x86-32$c))
                  (if (equal i j)
                      v
                    (mem$ci j x86-32$c)))))
\end{Verbatim}
}

It is well-known among ACL2 users that removing hypotheses of rules
can speed up the rewriter during proofs, or even make proofs possible
that might otherwise fail and require painful debugging when
hypotheses silently fail to prove.  The read-over-write theorem for
the model with abstract stobjs is as follows.

{\footnotesize
\begin{Verbatim}[commandchars=\\\{\}]
(defthm read-write
  (equal (memi i (!memi j v x86-32))
         (if (equal i j)
             (or v 0)
           (memi i x86-32))))
\end{Verbatim}
}

\noindent The use of records to represent the memory field made it possible to
eliminate the hypotheses, giving a stronger and cleaner theorem.

The use of abstract stobjs also benefits reasoning by avoiding certain proof
obligations for guard verification, by taking advantage of the fact
that the abstract stobj recognizer is preserved by single-threaded
code.  Consider the following definition, for the abstract stobj
\texttt{st} defined in Section~\ref{abstract-stobjs-example}.

{\footnotesize
\begin{verbatim}
(defun foo (st)
  (declare (xargs :stobjs st))
  (let ((st (update-misc 3 st)))
    (mv (misc st) st)))
\end{verbatim}
}

ACL2 accepts this definition without generating any proof obligations
for guard verification.  But without special treatment of stobj
recognizers, it would need to prove that \texttt{(STP ST)} implies
\texttt{(STP (UPDATE-MISC 3 ST))}.  This special treatment is afforded
concrete stobj recognizers as well, but would not be afforded
invariants defined on concrete stobjs.

\item {\em Layered Modeling Strategy:}\\
The use of abstract stobjs introduces a layer in the model.  As such,
the model becomes more manageable and robust.  For example, changes to
optimize the model for execution efficiency can be done on the
concrete layer. This would not affect the abstract layer, which
is used for reasoning, as long as the correspondence relation is
maintained.  A layered modeling strategy effectively eliminates the
need for a trade-off between reasoning and execution efficiency.

\end{itemize}

One might try to avoid abstract stobjs by defining two functions: a
``concrete'' one for execution that uses stobjs, and an ``abstract''
one for reasoning that does not.  For our Y86 example, a stobj-based
interpreter, {\tt run\$c}, could serve as our model and be used for
execution, while an auxiliary interpreter not using stobjs, {\tt
  run\$a}, could be used for proofs.  One might prove equivalence of
the two interpreters, using lemmas like some generated by
\texttt{defabsstobj} for a single step, and lifting to the run
functions using congruence-based reasoning.  The tricky bit could be
to explain exactly how this equivalence transfers a property proved
for {\tt (run\$a st\$a n)} to a property of {\tt (run\$c st\$c n')}.
Abstract stobjs avoid such challenges by providing a {\em single}
logical object with two representations.  Note also that the above
optimizations for guard checking and guard verification are not
available for a user-defined pair of models.

Note that it is possible to define more than one abstract stobj for a
single concrete stobj, which means that different representations of
the same stobj can be defined for different purposes.  We have not
exploited this fact, but we will find it interesting to learn of
applications that take advantage of it, so that different abstractions
can be used for different sets of proofs.

A traditional strength of ACL2 is its ability to provide both
efficient execution and effective reasoning.  Explicit support for
this combination includes the \texttt{mbe} and
~\texttt{defexec}~\cite{defexec} utilities for providing different
(but logically equal) code for execution and reasoning,\footnote{Both
  \texttt{mbe} and \texttt{defabsstobj} use \texttt{:LOGIC} and
  \texttt{:EXEC} keywords, but for \texttt{mbe} the functions are
  logically equal, while for \texttt{defabsstobj} exports they merely
  correspond, in the sense shown in Section~\ref{sec-abs-stobj-abs}.
  Single-threadedness seems crucial to us in maintaining a
  correspondence, but we have not explored extending the ideas of
  abstract stobjs to relax equality to correspondence without
  insisting on single-threadedness.} as well as
\texttt{defattach}~\cite{defattach-workshop}, which supports the
refinement of a constrained function by attaching an executable
function to it.  Single-threaded objects, and abstract stobjs in
particular, fit squarely into that tradition.  Such features
contribute to making ACL2 an industrial-strength system, up to the
tasks of modeling and proof for real processors.

\section*{Acknowledgements}

This material is based upon work supported by DARPA under contract
number N66001-10-2-4087 and by ForrestHunt, Inc.  We thank J Moore for
useful conversations; in particular, it was his suggestion that {\tt
  defabsstobj} print out missing events.  Sol Swords reported bugs in
earlier implementations of abstract stobjs, and we are grateful to him
for that and for helpful discussions, in particular pertaining to
congruent abstract stobjs and guard issues.  We also thank Robert Krug
for suggesting that generated events have rule-classes of
\texttt{nil}.  Finally, we thank the reviewers for their useful,
thorough reviews.

\bibliographystyle{eptcs}
\bibliography{goel-hunt-kaufmann}

\end{document}